\documentclass[preprintnumbers,amsmath,amssymb]{revtex4}

\usepackage{graphicx}
\usepackage{dcolumn}
\usepackage{bm}

\begin{document}

\preprint{}

{\bf \Large Quantum logic network for probabilistic cloning the quantum states }
\begin{center}
\begin{minipage}{15cm}
 {\large Ting Gao}\\
  {\it \small Department of Mathematics, Capital Normal University, Beijing 100037,
 China\\
  College of Mathematics and Information Science, Hebei Normal
University, Shijiazhuang
050016, China}\\[0.2cm]
{\large Fengli Yan }\\
{\it \small  Department of Physics, Hebei Normal University, Shijiazhuang 050016, China\\
 CCAST (World Laboratory), P.O. Box 8730, Beijing 100080, China}\\[0.2cm]
 {\large Zhixi Wang }\\
{\it \small  Department of Mathematics, Capital Normal University, Beijing 100037,
 China}\\[0.2cm]
{\small (Received 6 November 2003)\\[0.2cm]
We construct efficient quantum logic network  for probabilistic cloning the quantum states used in implemented
tasks for which cloning provides some enhancement in performance.}\\[0.2cm]
\end{minipage}
\end{center}
{\noindent \bf\large  I. INTRODUCTION  }\\[0.1cm]

Cloning is  a type of quantum information processing  tool. In 1982 Wootters and Zurek $^1$ and Dieks $^2$
independently  discovered that it is impossible to make perfect copies of an unknown quantum state. Since then
much effort has been put into developing optimal cloning processes $^{3-13}$. There are two main approaches to
quantum cloning. The first one invented by Bu\v{z}ek and Hillery relies on adding some ancillary quantum system
in a known state and unitarily evolving the resulting combined system, deterministically obtaining a pure state
with partial mixed density matrices (the clones) that are as close as possible to the original state $^3$. Duan
and Guo designed  the second kind of cloning procedure which is nondeterministic, consisting in adding an
ancilla, performing unitary operations and measurements, with a postselection of the measurement results
$^{13}$. The resulting clones are perfect, but the procedure only succeeds with a certain probability $p<1$,
which depends on the particular set of the states that we are trying to clone. Recently, Galv\~{a}o and Hardy
discuss how quantum information distribution implemented with different types of quantum cloning procedures can
improve the performance of  some quantum computation tasks $^{14}$. Unfortunately in the second example they
obtained the achievable efficiencies for probabilistic cloning the states by  a numerical search. Evidently the
numerical result is not an exact solution. Gao et al $^{15}$ provided exact achievable efficiencies for
probabilistically cloning the quantum states in Ref.14 used in implemented tasks for which cloning provides some
enhancement in performance. Clearly the quantum logic network for probabilistic cloning will be important in
realizing the cloning scheme in the experiment.
 Barenco et al showed that all unitary operations on arbitrarily
 many bits can be decomposed into the combinations of a set of one-bit quantum gates and two-bit Controlled-Not (CNOT)
 gates $^{16}$. In terms of only single-qubit gates,
two-qubit CNOT gates,  in this paper we present efficient quantum logic network for implementing probabilistic
cloning the quantum states in Ref.14.\\[0.1cm]

{\noindent \bf\large II. NOTATION }\\[0.1cm]

Some notations and symbols used throughout the paper are displayed here.
\begin{equation}\label{1}
 I=\left (
\begin{array} {cc}
1&0\\
0&1\\
\end{array}\right ); ~~~
X=\left (
\begin{array} {cc}
0&1\\
1&0\\
\end{array}\right ).
 \end{equation}
 For any unitary operator
 \begin{equation}\label{2}
 u=\left (
\begin{array} {cc}
u_{00}&u_{01}\\
u_{10}&u_{11}\\
\end{array}\right ),
\end{equation}
   define the 6-bit ($2^6$-dimensional) controlled operations $\Lambda_1(u)$, $\Lambda_2(u)$, $\Lambda_3(u)$,
 $\Lambda_4(u)$, $\Lambda_5(u)$, $\Lambda_6(u)$
as
 \begin{equation}\label{3}\begin{array}{l}
\Lambda_1(u)|x_1x_2\cdots x_6\rangle=(u^{x_2x_3x_4x_5x_6}|x_1\rangle)|x_2x_3x_4x_5x_6\rangle,\\
\Lambda_2(u)|x_1x_2\cdots x_6\rangle=|x_1\rangle(u^{x_1x_3x_4x_5x_6}|x_2\rangle)|x_3x_4x_5x_6\rangle,\\
\Lambda_3(u)|x_1x_2\cdots x_6\rangle=|x_1x_2\rangle(u^{x_1x_2x_4x_5x_6}|x_3\rangle)|x_4x_5x_6\rangle,\\
\Lambda_4(u)|x_1x_2\cdots x_6\rangle=|x_1x_2x_3\rangle(u^{x_1x_2x_3x_5x_6}|x_4\rangle)|x_5x_6\rangle,\\
\Lambda_5(u)|x_1x_2\cdots x_6\rangle=|x_1x_2x_3x_4\rangle(u^{x_1x_2x_3x_4x_6}|x_5\rangle)|x_6\rangle,\\
 \Lambda_6(u)|x_1x_2\cdots x_6\rangle=|x_1x_2x_3x_4x_5\rangle u^{x_1x_2x_3x_4x_5}|x_6\rangle,
 \end{array}\end{equation}
for all  $x_1, x_2, \cdots, x_6\in \{0, 1\}$. Here $x_{i_1}x_{i_2}x_{i_3}x_{i_4}x_{i_5}$ ($i_1, i_2, i_3,
i_4,i_5\in\{1,2,3,4,5,6\}$) in the exponent of $u$ means the product of the bits $x_{i_1}, x_{i_2}, x_{i_3},
x_{i_4}, x_{i_5}$. That is, the operator $u$ is applied to one qubit if other $5$ qubits are all equal to one,
and otherwise, nothing is done.  The $2^6\times 2^6$ matrices corresponding to $\Lambda_1(u)$, $\Lambda_2(u)$,
$\Lambda_3(u)$, $\Lambda_4(u)$, $\Lambda_5(u)$, $\Lambda _6(u)$ are
\begin{equation}\label{4}
\Lambda_1(u)=\left(%
\begin{array}{cccc}
  I_{31} &  &  &  \\
   & u_{00} & & u_{01} \\
   & & I_{31} & \\
   & u_{10} &  & u_{11}\\
\end{array}%
\right),~~
\Lambda_2(u)=\left(%
\begin{array}{cccc}
  I_{47} &  &  &  \\
   & u_{00} & & u_{01} \\
   & & I_{15} & \\
   & u_{10} &  & u_{11}\\
\end{array}%
\right),~~
\Lambda_3(u)=\left(%
\begin{array}{cccc}
  I_{55} &  &  &  \\
   & u_{00} & & u_{01} \\
   & & I_{7} & \\
   & u_{10} &  & u_{11}\\
\end{array}%
\right),
 \end{equation}
\begin{equation}\label{5}
\Lambda_4(u)=\left(%
\begin{array}{cccc}
  I_{59} &  &  &  \\
   & u_{00} & & u_{01} \\
   & & I_{3} & \\
   & u_{10} &  & u_{11}\\
\end{array}%
\right),~~~
\Lambda_5(u)=\left(%
\begin{array}{cccc}
  I_{61} &  &  &  \\
   & u_{00} & & u_{01} \\
   & & I_1 & \\
   & u_{10} &  & u_{11}\\
\end{array}%
\right),~~~
 \Lambda _6(u)=\left(
\begin{array} {ccc}
I_{62}&&\\
&u_{00}&u_{01}\\
&u_{10}&u_{11}
\end{array}\right)
\end{equation}
 (where $I_n$ denotes $n\times n$ identity matrix and the basis states are lexicographically ordered, i.e.,
 $|000000\rangle, |000001\rangle,
|000010\rangle,  \cdots, |111111\rangle$). $\Lambda_1(X)$, $\Lambda_2(X)$, $\Lambda_3(X)$, $\Lambda_4(X)$,
$\Lambda_5(X)$, $\Lambda_6(X)$ map  $|x_1x_2\cdots x_6\rangle$ to $|(x_2\wedge x_3\wedge x_4\wedge x_5\wedge
x_6)\oplus x_1\rangle|x_2x_3x_4x_5x_6\rangle$, $|x_1\rangle|(x_1\wedge x_3\wedge x_4\wedge x_5\wedge x_6)\oplus
x_2\rangle|x_3x_4x_5x_6\rangle$, $|x_1x_2\rangle|(x_1\wedge x_2\wedge x_4\wedge x_5\wedge x_6)\oplus
x_3\rangle|x_4x_5x_6\rangle$, $|x_1x_2x_3\rangle|(x_1\wedge x_2\wedge x_3\wedge x_5\wedge x_6)\oplus
x_4\rangle|x_5x_6\rangle$, $|x_1x_2x_3x_4\rangle|(x_1\wedge x_2\wedge x_3\wedge x_4\wedge x_6)\oplus
x_5\rangle|x_6\rangle$, $|x_1x_2x_3x_4x_5\rangle|(\wedge_{k=1}^5x_k)\oplus x_6\rangle$ respectively
($\wedge_{k=1}^5x_{i_k}$ stands for the AND of the Boolean variables {$x_{i_k}$}); that is, if  $5$ qubits are
all set to $|1\rangle$ then the other one qubit is flipped, otherwise the other one qubit is left alone.

Similarly, we define the ($n$+1)-bit ($2^{n+1}$-dimensional) controlled operation $C^n(u)$ as
 \begin{equation}\label{6}
 C^n(u)|x_1x_2\cdots x_n\rangle|y\rangle=|x_1x_2\cdots x_n\rangle u^{x_1x_2\cdots x_n}|y\rangle
 \end{equation}
for all  $x_1, x_2, \cdots, x_n, y\in \{0, 1\}$. Note that $C^0(u)$ is equated with $u$. The $2^{(n+1)}\times
2^{(n+1)}$ matrix corresponding to $C^n(u)$ is
\begin{equation}\label{7}
 \left (
\begin{array} {cccccccc}
1&&&&&\\
&1&&&&\\
&&\ddots&&&\\
&&&1&&\\
&&&&u_{00}&u_{01}\\
&&&&u_{10}&u_{11}
\end{array}\right ).
\end{equation}\\[0.1cm]

{\noindent \bf\large III. QUANTUM LOGIC NETWORK FOR PROBABILISTIC CLONING THE QUANTUM STATES}\\[0.1cm]

In Ref.14, the probabilistic cloning quantum states of a system $A$ consisting of two qubits are
\begin{equation}\label{8}\begin{array}{c}
|h_1\rangle=|h_{0010}\rangle\equiv\frac {1}{2}[|00\rangle+|01\rangle-|10\rangle+|11\rangle],\\
|h_2\rangle=|h_{0101}\rangle\equiv\frac
{1}{2}[|00\rangle-|01\rangle+|10\rangle-|11\rangle],\\
~~|h_3\rangle=|h_{1001}\rangle\equiv\frac {1}{2}[-|00\rangle+|01\rangle+|10\rangle-|11\rangle].
\end{array}
\end{equation}
Let $\gamma_1\equiv \gamma(|h_{0010}\rangle)$, $\gamma_2\equiv \gamma(|h_{0101}\rangle)$, $\gamma_3\equiv
\gamma(|h_{1001}\rangle)$ be the achievable efficiencies. In Ref.15 we discussed the maximum of the overall
probability (score) of success with the help of probabilistic cloning , defined the flag states and provided
exact achievable
 efficiencies
 \begin{equation}\label{9}
\gamma_1=\frac {1}{7}, ~~~~~~
 \gamma_2=\gamma_3=\frac {4}{7}
\end{equation}
 for probabilistic cloning the quantum states in Eq.(8).

Next we will give the efficient quantum logic network for  probabilistic cloning the quantum states in Eq.(8)
used in implemented tasks for which cloning provides some enhancement in performance.

From Eq.(9) of Ref.13 and the above  Eq.(9), we know that the unitary operation $U$ of probabilistic cloning the
quantum states in Eq.(8) should satisfy
\begin{equation}\label{10}
\left\{ \begin{array}{c}
U(|h_1\rangle|\Sigma\rangle|P_0\rangle)=\frac{1}{\sqrt{7}}|h_1\rangle|h_1\rangle|P^{(1)}\rangle+\sqrt{\frac{6}{7}}|\Phi^{(1)}\rangle,\\
U(|h_2\rangle|\Sigma\rangle|P_0\rangle)=\frac{2}{\sqrt{7}}|h_2\rangle|h_2\rangle|P^{(2)}\rangle+\sqrt{\frac{3}{7}}|\Phi^{(2)}\rangle,\\
U(|h_3\rangle|\Sigma\rangle|P_0\rangle)=\frac{2}{\sqrt{7}}|h_3\rangle|h_3\rangle|P^{(3)}\rangle+\sqrt{\frac{3}{7}}|\Phi^{(3)}\rangle.
\end{array}\right.
\end{equation}
Here $|\Sigma\rangle$ is the input state of an ancillary system $B$ being composed of two qubits; $|P_0\rangle$,
$|P^{(1)}\rangle$, $|P^{(2)}\rangle$ and $|P^{(3)}\rangle$ are 4 normalized states of the flag $P$; and
$|\Phi^{(1)}\rangle$, $|\Phi^{(2)}\rangle$ and  $|\Phi^{(3)}\rangle$ are 3 normalized states of the composite
system $ABP$.

By  Eqs.(10)-(11) provided by Duan and Guo in Ref.13, we may choose that
\begin{equation}\label{11}
\begin{array}{l}
|\Phi^{(1)}\rangle=|\Phi^{(1)}_{ABP}\rangle=\frac{1}{\sqrt{2}}|0000\rangle(|01\rangle+|10\rangle),\\
|\Phi^{(2)}\rangle=|\Phi^{(2)}_{ABP}\rangle=-|0000\rangle|01\rangle,\\
|\Phi^{(3)}\rangle=|\Phi^{(3)}_{ABP}\rangle=-|0000\rangle|10\rangle,\\
|P_0\rangle=|P^{(1)}\rangle=-|P^{(2)}\rangle=-|P^{(3)}\rangle=|00\rangle,\\
|\Sigma\rangle=|00\rangle.
\end{array}
\end{equation}
Here we emphasize  that the flag $P$ must consist of two qubits in order to satisfy the probabilistic cloning
condition (see Ref.13). Thus, on the basis
 $\{|000000\rangle, |000001\rangle,
|000010\rangle,  \cdots, |111111\rangle\}$, it is not difficult to prove that the unitary operation
\begin{eqnarray}\label{12}
U=&&P_{53,57}P_{49,53}P_{45,49}P_{37,45}P_{33,41}P_{29,37}P_{25,33}P_{21,33}P_{18,61}P_{17,29}P_{16,57}P_{15,53}P_{14,49}
P_{13,25}P_{12,45}P_{11,41}P_{10,37}P_{9,21}P_{8,33}\nonumber\\
&&P_{7,29}P_{6,25}P_{5,21} P_{4,21}
 \left(
\begin{array} {cc}
V&0\\
0&I_{46}\\
\end{array}\right)P_{11,12}P_{1,14}P_{7,15}P_{13,15}
P_{4,21}P_{5,21}P_{6,25}P_{7,29}P_{8,33}P_{9,21}P_{10,37}P_{11,41}P_{12,45}P_{13,25}\nonumber\\
&&P_{14,49}P_{15,53}
P_{16,57}P_{17,29}P_{18,61}P_{21,33}P_{25,33}P_{29,37}P_{33,41}P_{37,45}P_{45,49}P_{49,53}P_{53,57}
\end{eqnarray}
 satisfies Eq.(10).
 Here
\begin{equation}\label{13} V= \left(%
\begin{array}{cccccccccccccccccccccccccccccccc}
    \frac{1}{{\sqrt{2}}} & 0 & 0 & 0 & 0 & 0 & 0 & \frac{1}{2} & 0 & 0 & \frac{1}{2\,{\sqrt{3}}} & \frac{-1}
   {2\,{\sqrt{7}}} & \frac{1}{2\,{\sqrt{7}}} & \frac{-1}{4\,{\sqrt{7}}} & \frac{-1}{4\,{\sqrt{7}}} & \frac{1}
   {{\sqrt{42}}} & -{\sqrt{\frac{3}{70}}} & \frac{-{\sqrt{\frac{3}{70}}}}
   {2} \cr 0 & 0 & 0 & 0 & 0 & 0 & 0 & 0 & 0 & 0 & 0 & \frac{-{\sqrt{\frac{3}{7}}}}{2} & \frac{{\sqrt{\frac{3}{7}}}}
   {2} & 0 & {\sqrt{\frac{3}{7}}} & 0 & 2\,{\sqrt{\frac{2}{35}}} & \frac{-3}
   {{\sqrt{70}}} \cr 0 & 0 & 0 & 0 & 0 & 0 & 0 & 0 & 0 & 0 & 0 & \frac{-{\sqrt{\frac{3}{7}}}}{2} & \frac{{\sqrt{
         \frac{3}{7}}}}{2} & {\sqrt{\frac{3}{7}}} & 0 & 0 & 0 & {\sqrt{\frac{5}{14}}} \cr 0 & 0 & 0 & \frac{1}
   {{\sqrt{2}}} & 0 & 0 & 0 & 0 & 0 & 0 & 0 & \frac{1}{2\,{\sqrt{7}}} & \frac{-1}{2\,{\sqrt{7}}} & \frac{3}
   {4\,{\sqrt{7}}} & \frac{3}{4\,{\sqrt{7}}} & {\sqrt{\frac{3}{14}}} & -{\sqrt{\frac{3}{70}}} & \frac{-{\sqrt{
          \frac{3}{70}}}}{2} \cr 0 & 0 & 0 & 0 & \frac{1}{{\sqrt{2}}} & 0 & 0 & 0 & -\frac{1}{2}
      & 0 & 0 & 0 & 0 & \frac{-3}{4\,{\sqrt{7}}} & \frac{1}{4\,{\sqrt{7}}} & \frac{{\sqrt{\frac{3}{14}}}}{2} &
    \frac{{\sqrt{\frac{3}{70}}}}{2} & \frac{3\,{\sqrt{\frac{3}{70}}}}{2} \cr 0 & 0 & 0 & 0 & \frac{1}
   {{\sqrt{2}}} & 0 & 0 & 0 & \frac{1}{2} & 0 & 0 & 0 & 0 & \frac{3}{4\,{\sqrt{7}}} & \frac{-1}
   {4\,{\sqrt{7}}} & \frac{-{\sqrt{\frac{3}{14}}}}{2} & \frac{-{\sqrt{\frac{3}{70}}}}{2} & \frac{-3\,
     {\sqrt{\frac{3}{70}}}}{2} \cr 0 & 0 & 0 & -\frac{1}{{\sqrt{2}}}
      & 0 & 0 & 0 & 0 & 0 & 0 & 0 & \frac{1}{2\,{\sqrt{7}}} & \frac{-1}{2\,{\sqrt{7}}} & \frac{3}
   {4\,{\sqrt{7}}} & \frac{3}{4\,{\sqrt{7}}} & {\sqrt{\frac{3}{14}}} & -{\sqrt{\frac{3}{70}}} & \frac{-{\sqrt{
          \frac{3}{70}}}}{2} \cr -\frac{1}{{\sqrt{2}}}  & 0 & 0 & 0 & 0 & 0 & 0 & \frac{1}
   {2} & 0 & 0 & \frac{1}{2\,{\sqrt{3}}} & \frac{-1}{2\,{\sqrt{7}}} & \frac{1}{2\,{\sqrt{7}}} & \frac{-1}
   {4\,{\sqrt{7}}} & \frac{-1}{4\,{\sqrt{7}}} & \frac{1}{{\sqrt{42}}} & -{\sqrt{\frac{3}{70}}} & \frac{-{\sqrt{
          \frac{3}{70}}}}{2} \cr 0 & \frac{1}{{\sqrt{2}}} & 0 & 0 & 0 & 0 & 0 & 0 & 0 & \frac{1}
   {2} & 0 & 0 & 0 & \frac{1}{4\,{\sqrt{7}}} & \frac{-3}{4\,{\sqrt{7}}} & \frac{{\sqrt{\frac{3}{14}}}}{2} &
    \frac{3\,{\sqrt{\frac{3}{70}}}}{2} & \frac{-{\sqrt{\frac{3}{70}}}}{2} \cr 0 & \frac{1}
   {{\sqrt{2}}} & 0 & 0 & 0 & 0 & 0 & 0 & 0 & -\frac{1}{2}  & 0 & 0 & 0 & \frac{-1}
   {4\,{\sqrt{7}}} & \frac{3}{4\,{\sqrt{7}}} & \frac{-{\sqrt{\frac{3}{14}}}}{2} & \frac{-3\,{\sqrt{\frac{3}{70}}}}
   {2} & \frac{{\sqrt{\frac{3}{70}}}}{2} \cr 0 & 0 & 0 & 0 & 0 & 0 & 0 & 0 & \frac{1}{2} & 0 & 0 & \frac{1}
   {2} & \frac{1}{2} & \frac{-3}{4\,{\sqrt{7}}} & \frac{1}{4\,{\sqrt{7}}} & \frac{{\sqrt{\frac{3}{14}}}}{2} &
    \frac{{\sqrt{\frac{3}{70}}}}{2} & \frac{3\,{\sqrt{\frac{3}{70}}}}{2} \cr 0 & 0 & 0 & 0 & 0 & 0 & \frac{1}
   {{\sqrt{2}}} & 0 & 0 & -\frac{1}{2}  & 0 & 0 & 0 & \frac{1}{4\,{\sqrt{7}}} & \frac{-3}
   {4\,{\sqrt{7}}} & \frac{{\sqrt{\frac{3}{14}}}}{2} & \frac{3\,{\sqrt{\frac{3}{70}}}}{2} & \frac{-{\sqrt{\frac{3}
         {70}}}}{2} \cr 0 & 0 & \frac{1}{{\sqrt{2}}} & 0 & 0 & 0 & 0 & -\frac{1}{2}  & 0 & 0 &
    \frac{1}{2\,{\sqrt{3}}} & \frac{-1}{2\,{\sqrt{7}}} & \frac{1}{2\,{\sqrt{7}}} & \frac{-1}{4\,{\sqrt{7}}} &
    \frac{-1}{4\,{\sqrt{7}}} & \frac{1}{{\sqrt{42}}} & -{\sqrt{\frac{3}{70}}} & \frac{-{\sqrt{\frac{3}{70}}}}
   {2} \cr 0 & 0 & \frac{1}{{\sqrt{2}}} & 0 & 0 & 0 & 0 & \frac{1}{2} & 0 & 0 & \frac{-1}{2\,{\sqrt{3}}} & \frac{1}
   {2\,{\sqrt{7}}} & \frac{-1}{2\,{\sqrt{7}}} & \frac{1}{4\,{\sqrt{7}}} & \frac{1}{4\,{\sqrt{7}}} & -\frac{1}
     {{\sqrt{42}}}  & {\sqrt{\frac{3}{70}}} & \frac{{\sqrt{\frac{3}{70}}}}
   {2} \cr 0 & 0 & 0 & 0 & 0 & 0 & 0 & 0 & -\frac{1}{2}  & 0 & 0 & \frac{1}{2} & \frac{1}{2} &
    \frac{3}{4\,{\sqrt{7}}} & \frac{-1}{4\,{\sqrt{7}}} & \frac{-{\sqrt{\frac{3}{14}}}}{2} & \frac{-{\sqrt{\frac{3}
         {70}}}}{2} & \frac{-3\,{\sqrt{\frac{3}{70}}}}{2} \cr 0 & 0 & 0 & 0 & 0 & 0 & \frac{1}
   {{\sqrt{2}}} & 0 & 0 & \frac{1}{2} & 0 & 0 & 0 & \frac{-1}{4\,{\sqrt{7}}} & \frac{3}{4\,{\sqrt{7}}} & \frac{-
       {\sqrt{\frac{3}{14}}}}{2} & \frac{-3\,{\sqrt{\frac{3}{70}}}}{2} & \frac{{\sqrt{\frac{3}{70}}}}
   {2} \cr 0 & 0 & 0 & 0 & 0 & \frac{1}{{\sqrt{2}}} & 0 & 0 & 0 & 0 & \frac{1}{{\sqrt{3}}} & \frac{1}
   {2\,{\sqrt{7}}} & \frac{-1}{2\,{\sqrt{7}}} & \frac{1}{4\,{\sqrt{7}}} & \frac{1}{4\,{\sqrt{7}}} & -\frac{1}
     {{\sqrt{42}}}  & {\sqrt{\frac{3}{70}}} & \frac{{\sqrt{\frac{3}{70}}}}{2} \cr 0 & 0 & 0 & 0 & 0 &
    \frac{1}{{\sqrt{2}}} & 0 & 0 & 0 & 0 & -\frac{1}{{\sqrt{3}}}  & \frac{-1}{2\,{\sqrt{7}}} &
    \frac{1}{2\,{\sqrt{7}}} & \frac{-1}{4\,{\sqrt{7}}} & \frac{-1}{4\,{\sqrt{7}}} & \frac{1}{{\sqrt{42}}} & -{
      \sqrt{\frac{3}{70}}} & \frac{-{\sqrt{\frac{3}{70}}}}{2} \cr
 \end{array}%
\right),
\end{equation} $P_{i,j}$ denote the following $64\times 64$
the matrix in which the two off-diagonal 1's are in the i,j and j,i positions and all unspecified entries are 0
\begin{equation}\label{}P_{i,j}=~~~~~~
 \bordermatrix{&  &  & &$column$~i & & & & \mbox {column}~j & & &\cr
  &1 &  &  &  &  &  &  &  &  &  & \cr
  & & \ddots &  &  &  &  &  &  &  &  & \cr
  & &  & 1 &  &  &  &  &  &  &  & \cr
  $row$~i &  &  &  & 0 &  &  &  & 1 &  &  & \cr
  &  &  &  &  & 1 &  &  &  &  &  & \cr
  &  &   &  &  &  & \ddots &  &  &  &  &  &  & \cr
 &   &  &  &  &  &  & 1 &  & \cr
$row$~j  &   &  &  & 1 &  &  &  & 0 &  &  & \cr
 &   &  &  &  &  &  &  &  &  1 &  & \cr
  &   &  &  &  &  &  &  &  &  &\ddots & \cr
 &   &  &  &  &  &  &  &  &  &  &1\cr}.
\end{equation}

$U$ may be written

\begin{eqnarray}U=&&P_{53,57}P_{49,53}P_{45,49}P_{37,45}P_{33,41}P_{29,37}P_{25,33}P_{21,33}P_{18,61}P_{17,29}P_{16,57}
P_{15,53}P_{14,49}P_{13,25}P_{12,45}P_{11,41}P_{10,37}P_{9,21}P_{8,33}\nonumber\\
&&P_{7,29}P_{6,25}P_{5,21} P_{4,21}\left(%
\begin{array}{cc}
  \frac{1}{\sqrt{2}} & -\frac{1}{\sqrt{2}} \\
  -\frac{1}{\sqrt{2}} & -\frac{1}{\sqrt{2}} \\
\end{array}%
\right)_{1,8}P_{2,9}\left(%
\begin{array}{cc}
  \frac{1}{\sqrt{2}} & \frac{1}{\sqrt{2}} \\
  \frac{1}{\sqrt{2}} & -\frac{1}{\sqrt{2}} \\
\end{array}%
\right)_{2,10}P_{3,13}\left(%
\begin{array}{cc}
  \frac{1}{\sqrt{2}} & \frac{1}{\sqrt{2}} \\
  \frac{1}{\sqrt{2}} & -\frac{1}{\sqrt{2}} \\
\end{array}%
\right)_{3,14}\left(%
\begin{array}{cc}
  \frac{1}{\sqrt{2}} & -\frac{1}{\sqrt{2}} \\
  -\frac{1}{\sqrt{2}} & -\frac{1}{\sqrt{2}} \\
\end{array}%
\right)_{4,7}\nonumber\\
&&\left(%
\begin{array}{cc}
  \frac{1}{\sqrt{2}} & \frac{1}{\sqrt{2}} \\
  \frac{1}{\sqrt{2}} & -\frac{1}{\sqrt{2}} \\
\end{array}%
\right)_{5,6}P_{6,17}\left(%
\begin{array}{cc}
  \frac{1}{\sqrt{2}} & \frac{1}{\sqrt{2}} \\
  \frac{1}{\sqrt{2}} & -\frac{1}{\sqrt{2}} \\
\end{array}%
\right)_{6,18}P_{7,12}\left(%
\begin{array}{cc}
  \frac{1}{\sqrt{2}} & \frac{1}{\sqrt{2}} \\
  \frac{1}{\sqrt{2}} & -\frac{1}{\sqrt{2}} \\
\end{array}%
\right)_{7,16}\left(%
\begin{array}{cc}
  -\frac{1}{\sqrt{2}} & -\frac{1}{\sqrt{2}} \\
  -\frac{1}{\sqrt{2}} & \frac{1}{\sqrt{2}} \\
\end{array}%
\right)_{8,14}P_{9,11}\left(%
\begin{array}{cc}
  \frac{1}{\sqrt{2}} & -\frac{1}{\sqrt{2}} \\
  -\frac{1}{\sqrt{2}} & -\frac{1}{\sqrt{2}} \\
\end{array}%
\right)_{9,15}\nonumber\\
&&
\left(%
\begin{array}{cc}
  \frac{1}{\sqrt{2}} & -\frac{1}{\sqrt{2}} \\
  -\frac{1}{\sqrt{2}} & -\frac{1}{\sqrt{2}} \\
\end{array}%
\right)_{9,17}\left(%
\begin{array}{cc}
  \frac{1}{\sqrt{2}} & -\frac{1}{\sqrt{2}} \\
  -\frac{1}{\sqrt{2}} & -\frac{1}{\sqrt{2}} \\
\end{array}%
\right)_{10,16}P_{11,14}\left(%
\begin{array}{cc}
  \frac{1}{\sqrt{3}} & \sqrt{\frac{2}{3}}\\
  \sqrt{\frac{2}{3}} & - \frac{1}{\sqrt{3}}\\
\end{array}%
\right)_{11,18}\left(%
\begin{array}{cc}
  -\sqrt{\frac{2}{5}} & -\sqrt{\frac{3}{5}} \\
  -\sqrt{\frac{3}{5}} & \sqrt{\frac{2}{5}} \\
\end{array}%
\right)_{12,13}\left(%
\begin{array}{cc}
  \frac{\sqrt{\frac{5}{2}}}{2} & -\frac{\sqrt{\frac{3}{2}}}{2} \\
  -\frac{\sqrt{\frac{3}{2}}}{2}& -\frac{\sqrt{\frac{5}{2}}}{2} \\
\end{array}%
\right)_{12,14}\nonumber\\
&&\left(%
\begin{array}{cc}
  \frac{2}{\sqrt{11}} & -\sqrt{\frac{7}{11}} \\
  -\sqrt{\frac{7}{11}} & -\frac{2}{\sqrt{11}}\\
\end{array}%
\right)_{12,15}\left(%
\begin{array}{cc}
  \sqrt{\frac{11}{14}} & -\sqrt{\frac{3}{14}} \\
  -\sqrt{\frac{3}{14}} & -\sqrt{\frac{11}{14}} \\
\end{array}%
\right)_{12,18}P_{13,15}\left(%
\begin{array}{cc}
  2\sqrt{\frac{2}{11}} & -\sqrt{\frac{3}{11}} \\
  -\sqrt{\frac{3}{11}} & -2\sqrt{\frac{2}{11}} \\
\end{array}%
\right)_{13,18}\left(%
\begin{array}{cc}
  \frac{3}{\sqrt{205}} & \frac{14}{\sqrt{205}} \\
  \frac{14}{\sqrt{205}} & - \frac{3}{\sqrt{205}} \\
\end{array}%
\right)_{14,15}\nonumber\\
&&\left(%
\begin{array}{cc}
  \sqrt{\frac{123}{131}} & -2\sqrt{\frac{2}{131}} \\
  -2\sqrt{\frac{2}{131}} & -\sqrt{\frac{123}{131}}\\
\end{array}%
\right)_{14,16}\left(%
\begin{array}{cc}
  \sqrt{\frac{131}{203}} & 6\sqrt{\frac{2}{203}} \\
  6\sqrt{\frac{2}{203}} & -\sqrt{\frac{131}{203}}\\
\end{array}%
\right)_{14,17}\left(%
\begin{array}{cc}
  \frac{\sqrt{\frac{29}{2}}}{4} & -\frac{\sqrt{\frac{3}{2}}}{4} \\
  -\frac{\sqrt{\frac{3}{2}}}{4} & -\frac{\sqrt{\frac{29}{2}}}{4} \\
\end{array}%
\right)_{14,18}\left(%
\begin{array}{cc}
  -5\sqrt{\frac{131}{5453}} & -33\sqrt{\frac{2}{5453}} \\
  -33\sqrt{\frac{2}{5453}} & 5\sqrt{\frac{131}{5453}} \\
\end{array}%
\right)_{15,16}\nonumber\\
&&\left(%
\begin{array}{cc}
  \sqrt{\frac{1653}{1703}} & 5\sqrt{\frac{2}{1703}} \\
  5\sqrt{\frac{2}{1703}} & -\sqrt{\frac{1653}{1703}}\\
\end{array}%
\right)_{15,17}\left(%
\begin{array}{cc}
  \sqrt{\frac{26}{29}} & \sqrt{\frac{3}{29}} \\
  \sqrt{\frac{3}{29}} & -\sqrt{\frac{26}{29}} \\
\end{array}%
\right)_{15,18}\left(%
\begin{array}{cc}
  \sqrt{\frac{13}{38}} & -\frac{5}{\sqrt{38}} \\
  -\frac{5}{\sqrt{38}} & -\sqrt{\frac{13}{38}} \\
\end{array}%
\right)_{16,17}\left(%
\begin{array}{cc}
  \frac{3}{\sqrt{13}} & \frac{2}{\sqrt{13}} \\
  \frac{2}{\sqrt{13}} & -\frac{3}{\sqrt{13}} \\
\end{array}%
\right)_{16,18}\left(%
\begin{array}{cc}
  -\frac{1}{\sqrt{5}} & \frac{2}{\sqrt{5}} \\
  \frac{2}{\sqrt{5}} & \frac{1}{\sqrt{5}} \\
\end{array}%
\right)_{17,18}\nonumber\\
&&P_{11,12}P_{1,14}P_{7,15}P_{13,15}
P_{4,21}P_{5,21}P_{6,25}P_{7,29}P_{8,33}P_{9,21}P_{10,37}P_{11,41}P_{12,45}P_{13,25}P_{14,49}P_{15,53}
P_{16,57}P_{17,29}P_{18,61}P_{21,33}\nonumber\\
&&P_{25,33}P_{29,37}P_{33,41}P_{37,45}P_{45,49}P_{49,53}P_{53,57},
\end{eqnarray}
 where  the matrices
$$\left (
\begin{array} {cc}
u_{00}&u_{01}\\
u_{10}&u_{11}\\
\end{array}\right )_{i,j}=~~~~~~
 \bordermatrix{&  &  & &$column$~i & & & & \mbox {column}~j & & &\cr
  &1 &  &  &  &  &  &  &  &  &  & \cr
  & & \ddots &  &  &  &  &  &  &  &  & \cr
  & &  & 1 &  &  &  &  &  &  &  & \cr
  $row$~i &  &  &  & u_{00} &  &  &  & u_{01} &  &  & \cr
  &  &  &  &  & 1 &  &  &  &  &  & \cr
  &  &   &  &  &  & \ddots &  &  &  &  &  &  & \cr
 &   &  &  &  &  &  & 1 &  & \cr
$row$~j  &   &  &  & u_{10} &  &  &  & u_{11} &  &  & \cr
 &   &  &  &  &  &  &  &  &  1 &  & \cr
  &   &  &  &  &  &  &  &  &  &\ddots & \cr
 &   &  &  &  &  &  &  &  &  &  &1\cr}$$
are $64\times 64$ two-level unitary matrices.

$U$ can be expressed as

 $ U=(I\otimes I\otimes X\otimes I\otimes X\otimes X)\Lambda_4(X)(I\otimes I\otimes
X\otimes X\otimes I\otimes I)\Lambda_3(X)(I\otimes X\otimes I\otimes X\otimes I\otimes I)\Lambda_4(X)(I\otimes
I\otimes I\otimes X\otimes I\otimes I)\Lambda_3(X)(I\otimes X\otimes X\otimes I\otimes I\otimes
I)\Lambda_2(X)(I\otimes X\otimes X\otimes I\otimes I\otimes I)\Lambda_3(X)(I\otimes I\otimes I\otimes X\otimes
I\otimes I)\Lambda_4(X)\Lambda_3(X)(I\otimes I\otimes I\otimes X\otimes I\otimes I)\Lambda_3(X)(X\otimes
X\otimes I\otimes X\otimes I\otimes I)\Lambda_3(X)(I\otimes I\otimes X\otimes I\otimes I\otimes
I)\Lambda_2(X)(X\otimes X\otimes I\otimes I\otimes I\otimes I)\Lambda_1(X)(X\otimes X\otimes I\otimes I\otimes
I\otimes I)\Lambda_2(X)(I\otimes I\otimes X\otimes I\otimes I\otimes I)\Lambda_3(X)(I\otimes I\otimes I\otimes
X\otimes I\otimes I)\Lambda_3(X)(X\otimes I\otimes X\otimes I\otimes I\otimes
I)\Lambda_1(X)\Lambda_2(X)\Lambda_1(X)(X\otimes I\otimes X\otimes I\otimes I\otimes I)\Lambda_3(X)(I\otimes
I\otimes X\otimes X\otimes I\otimes I)\Lambda_4(X)(X\otimes I\otimes I\otimes X\otimes I\otimes
I)\Lambda_1(X)\Lambda_2(X)\Lambda_1(X)(X\otimes I\otimes I\otimes X\otimes I\otimes I)\Lambda_4(X)(I\otimes
I\otimes I\otimes X\otimes I\otimes X)\Lambda_6(X)(X\otimes I\otimes I\otimes I\otimes I\otimes
X)\Lambda_1(X)(I\otimes I\otimes X\otimes I\otimes I\otimes I)\Lambda_3(X)(I\otimes I\otimes I\otimes X\otimes
I\otimes I)\Lambda_4(X)(I\otimes I\otimes I\otimes X\otimes I\otimes I)\Lambda_3(X)(I\otimes I\otimes X\otimes
I\otimes I\otimes I)\Lambda_1(X)(X\otimes I\otimes I\otimes I\otimes I\otimes X)\Lambda_6(X)(I\otimes I\otimes
X\otimes I\otimes I\otimes X)\Lambda_3(X)(I\otimes I\otimes I\otimes X\otimes I\otimes I)\Lambda_4(X)(I\otimes
I\otimes I\otimes X\otimes I\otimes I)\Lambda_3(X)(I\otimes X\otimes I\otimes X\otimes X\otimes
X)\Lambda_6(X)(I\otimes I\otimes I\otimes I\otimes I\otimes X)\Lambda_5(X)(I\otimes I\otimes I\otimes I\otimes
X\otimes I)\Lambda_4(X)(I\otimes X\otimes I\otimes X\otimes I\otimes I)\Lambda_2(X)(X\otimes I\otimes I\otimes
I\otimes I\otimes I)\Lambda_1(X)(X\otimes I\otimes I\otimes I\otimes I\otimes I)\Lambda_2(X) (I\otimes X\otimes
I\otimes X\otimes I\otimes I)\Lambda_4(X)(I\otimes I\otimes I\otimes I\otimes X\otimes I)\Lambda_5(X)(I\otimes
I\otimes I\otimes I\otimes I\otimes X)\Lambda_6(X)(I\otimes I\otimes I\otimes I\otimes I\otimes
X)\Lambda_3(X)(I\otimes I\otimes X\otimes I\otimes I\otimes I)\Lambda_5(X)(I\otimes X\otimes I\otimes I\otimes
X\otimes I)\Lambda_2(X)(X\otimes I\otimes I\otimes I\otimes I\otimes I)\Lambda_1(X)(X\otimes I\otimes I\otimes
I\otimes I\otimes I)\Lambda_2(X)(I\otimes X\otimes I\otimes I\otimes X\otimes I)\Lambda_5(X)(I\otimes I\otimes
X\otimes I\otimes I\otimes I)\Lambda_3(X)(I\otimes I\otimes I\otimes I\otimes X\otimes X)\Lambda_6(X) (I\otimes
I\otimes I\otimes I\otimes I\otimes X)\Lambda_4(X)(I\otimes X\otimes I\otimes X\otimes I\otimes
I)\Lambda_2(X)\Lambda_3(X)(X\otimes I\otimes X\otimes I\otimes I\otimes I)\Lambda_1(X)(X\otimes I\otimes
X\otimes I\otimes I\otimes I)\Lambda_3(X)\Lambda_2(X)(I\otimes X\otimes I\otimes X\otimes I\otimes
I)\Lambda_4(X)(I\otimes I\otimes I\otimes I\otimes I\otimes X)\Lambda_6(X)(I\otimes X\otimes I\otimes I\otimes
I\otimes X)\Lambda_2(X)\Lambda_4(X)\Lambda_2(X)(I\otimes X\otimes I\otimes X\otimes X\otimes X)\Lambda_6(X)
(I\otimes I\otimes I\otimes I\otimes I\otimes X)\Lambda_5(X)(I\otimes I\otimes I\otimes X\otimes X\otimes I)
\Lambda_4(X)(X\otimes I\otimes I\otimes I\otimes I\otimes I)\Lambda_1(X)(X\otimes I\otimes I\otimes I\otimes
I\otimes I)\Lambda_4(X)(I\otimes I\otimes I\otimes X\otimes X\otimes I)\Lambda_5(X)(I\otimes I\otimes I\otimes
I\otimes I\otimes X)\Lambda_6(X)(X\otimes I\otimes I\otimes I\otimes I\otimes
X)\Lambda_1(X)\Lambda_5(X)\Lambda_1(X)(X\otimes I\otimes I\otimes I\otimes X\otimes X)\Lambda_6(X)(I\otimes
I\otimes I\otimes X\otimes I\otimes X)\Lambda_4(X)(X\otimes I\otimes I\otimes I\otimes I\otimes I)\Lambda_1(X)
 \Lambda_3(X) \Lambda_1(X)(X\otimes I\otimes I\otimes I\otimes I\otimes I)\Lambda_4(X)(I\otimes I\otimes I\otimes
X\otimes I\otimes X)\Lambda_6(X)(I\otimes X\otimes I\otimes I\otimes I\otimes X)\Lambda_2(X)(I\otimes I\otimes
I\otimes X\otimes I\otimes I)\Lambda_4(X)\Lambda_3(X)\Lambda_4(X)(I\otimes I\otimes I\otimes X\otimes I\otimes
I)\Lambda_2(X)(I\otimes X\otimes X\otimes X\otimes X\otimes X)\Lambda_6(X)(I\otimes I\otimes I\otimes I\otimes
I\otimes X)\Lambda_5(X)(X\otimes I\otimes I\otimes I\otimes X\otimes
I)\Lambda_1(X)\Lambda_4(X)\Lambda_1(X)(X\otimes I\otimes I\otimes I\otimes X\otimes I)\Lambda_5(X)(I\otimes
I\otimes I\otimes I\otimes I\otimes X)\Lambda_6(X)(I\otimes I\otimes X\otimes I\otimes I\otimes X)\Lambda_3(X)
\Lambda_5(X)(I\otimes X\otimes I\otimes I\otimes X\otimes I)\Lambda_2(X)(I\otimes X\otimes I\otimes I\otimes
X\otimes I)\Lambda_5(X)\Lambda_3(X)(I\otimes I\otimes I\otimes I\otimes X\otimes X)\Lambda_3(X)\Lambda_6(X)
(I\otimes I\otimes I\otimes I\otimes I\otimes X)\Lambda_4(X)(I\otimes X\otimes I\otimes X\otimes I\otimes I)
\Lambda_2(X)(I\otimes X\otimes I\otimes X\otimes I\otimes I)\Lambda_4(X)(I\otimes I\otimes I\otimes I\otimes
I\otimes X)\Lambda_6(X)\Lambda_3(X)(I\otimes X\otimes X\otimes I\otimes I\otimes X)\Lambda_2(X)(I\otimes
X\otimes I\otimes X\otimes X\otimes X)\Lambda_6(X)(I\otimes I\otimes I\otimes X\otimes I\otimes X)\Lambda_4(X)
\Lambda_5(X)(I\otimes X\otimes I\otimes I\otimes X\otimes I)\Lambda_2(X)(I\otimes X\otimes I\otimes I\otimes
X\otimes I)\Lambda_5(X)\Lambda_4(X)(I\otimes I\otimes I\otimes X\otimes I\otimes X)\Lambda_6(X)$

$(I\otimes I\otimes I\otimes I\otimes X\otimes I)\Lambda_6(X)(I\otimes I\otimes I\otimes I\otimes X\otimes
I)\Lambda_5(X)(I\otimes I\otimes I\otimes X\otimes I\otimes I)\Lambda_4(v_1)(I\otimes I\otimes I\otimes X\otimes
I\otimes I)\Lambda_5(X)(I\otimes I\otimes I\otimes I\otimes X\otimes I)\Lambda_6(X)(I\otimes I\otimes X\otimes
I\otimes I\otimes I)\Lambda_3(X)\Lambda_6(X)\Lambda_3(X)\Lambda_3(v_2)(I\otimes I\otimes X\otimes X\otimes
X\otimes X)\Lambda_4(X)\Lambda_5(X)(I\otimes I\otimes X\otimes I\otimes X\otimes X)\Lambda_6(v_2)(I\otimes
I\otimes I\otimes I\otimes I\otimes X)\Lambda_3(X)(I\otimes I\otimes X\otimes I\otimes X\otimes
I)\Lambda_5(X)\Lambda_4(X)(I\otimes I\otimes I\otimes X\otimes I\otimes X)\Lambda_6(X)(I\otimes I\otimes
I\otimes X\otimes I\otimes X)\Lambda_4(v_1)(I\otimes I\otimes I\otimes X\otimes I\otimes X)\Lambda_6(X)(I\otimes
I\otimes I\otimes X\otimes X\otimes I)\Lambda_6(v_2)\Lambda_4(X)(I\otimes X\otimes I\otimes X\otimes I\otimes
I)\Lambda_2(X)\Lambda_6(X)\Lambda_2(X)\Lambda_2(v_2)(I\otimes X\otimes I\otimes X\otimes I\otimes
I)\Lambda_4(X)(I\otimes I\otimes I\otimes I\otimes X\otimes I)\Lambda_6(X)\Lambda_4(X)(I\otimes I\otimes
X\otimes X\otimes I\otimes I)\Lambda_3(X)(I\otimes I\otimes X\otimes X\otimes I\otimes I)\Lambda_4(X)(I\otimes
I\otimes X\otimes I\otimes I\otimes I)\Lambda_3(v_2)(I\otimes I\otimes X\otimes I\otimes I\otimes
I)\Lambda_6(X)\Lambda_5(X)(I\otimes I\otimes X\otimes I\otimes X\otimes I)\Lambda_3(-v_2)(I\otimes I\otimes
X\otimes I\otimes X\otimes I)\Lambda_5(X)(I\otimes I\otimes X\otimes I\otimes I\otimes X)\Lambda_4(v_1)(I\otimes
I\otimes I\otimes X\otimes I\otimes I)\Lambda_5(X)(I\otimes X\otimes I\otimes I\otimes X\otimes
I)\Lambda_2(X)\Lambda_3(-v_2)\Lambda_2(X)(I\otimes X\otimes I\otimes I\otimes X\otimes I)\Lambda_5(X)(I\otimes
I\otimes I\otimes X\otimes I\otimes I)\Lambda_4(v_1)(I\otimes I\otimes I\otimes X\otimes I\otimes
I)\Lambda_5(X)\Lambda_6(X)\Lambda_5(X)(I\otimes I\otimes I\otimes X\otimes X\otimes I)\Lambda_4(X)(I\otimes
I\otimes I\otimes X\otimes I\otimes I)\Lambda_3(X)(I\otimes X\otimes X\otimes I\otimes I\otimes
I)\Lambda_2(v_3)(I\otimes X\otimes X\otimes I\otimes I\otimes I)\Lambda_3(X)(I\otimes I\otimes I\otimes I\otimes
X\otimes I)\Lambda_5(X)(I\otimes I\otimes I\otimes I\otimes I\otimes X)\Lambda_5(X)(I\otimes I\otimes I\otimes
X\otimes X\otimes I)\Lambda_4(v_4)(I\otimes I\otimes I\otimes X\otimes X\otimes I)\Lambda_5(X)(I\otimes I\otimes
I\otimes I\otimes I\otimes X)\Lambda_6(X)\Lambda_5(X)(I\otimes I\otimes I\otimes X\otimes X\otimes
I)\Lambda_4(v_5)(I\otimes I\otimes I\otimes X\otimes X\otimes I)\Lambda_5(X)\Lambda_6(X)(I\otimes I\otimes
I\otimes X\otimes I\otimes X)\Lambda_4(v_6)(I\otimes I\otimes I\otimes X\otimes I\otimes
X)\Lambda_6(X)\Lambda_3(X)(I\otimes I\otimes X\otimes I\otimes I\otimes I)\Lambda_5(X)(I\otimes X\otimes
I\otimes I\otimes X\otimes I)\Lambda_2(v_7)(I\otimes X\otimes I\otimes I\otimes X\otimes I)\Lambda_5(X)(I\otimes
I\otimes X\otimes I\otimes I\otimes I)\Lambda_3(X)(I\otimes I\otimes I\otimes X\otimes I\otimes
X)\Lambda_5(X)(I\otimes I\otimes I\otimes I\otimes X\otimes X)\Lambda_6(X)\Lambda_3(X)(I\otimes I\otimes
X\otimes I\otimes I\otimes I)\Lambda_4(X)(I\otimes X\otimes I\otimes X\otimes I\otimes I)\Lambda_2(v_8)(I\otimes
X\otimes I\otimes X\otimes I\otimes I)\Lambda_4(X)(I\otimes I\otimes X\otimes I\otimes I\otimes
I)\Lambda_3(X)(I\otimes I\otimes I\otimes I\otimes X\otimes X)\Lambda_5(v_9)(I\otimes I\otimes I\otimes I\otimes
X\otimes X)\Lambda_6(X)(I\otimes I\otimes I\otimes I\otimes X\otimes I)\Lambda_5(v_{10})(I\otimes I\otimes
I\otimes I\otimes X\otimes I)\Lambda_3(X)(I\otimes I\otimes X\otimes I\otimes I\otimes I)\Lambda_4(X)(I\otimes
X\otimes I\otimes X\otimes I\otimes I)\Lambda_2(X)\Lambda_6(v_{11})\Lambda_2(X)\Lambda_2(v_{12})(I\otimes
X\otimes I\otimes X\otimes I\otimes I)\Lambda_4(X)(I\otimes I\otimes X\otimes I\otimes I\otimes
I)\Lambda_3(X)(I\otimes I\otimes I\otimes I\otimes X\otimes I)\Lambda_6(v_{13})(I\otimes I\otimes I\otimes
I\otimes I\otimes X)\Lambda_3(X)(I\otimes I\otimes X\otimes I\otimes I\otimes I)\Lambda_4(X)(I\otimes X\otimes
I\otimes X\otimes I\otimes I)\Lambda_2(X)\Lambda_5(v_{14})\Lambda_5(X)(I\otimes I\otimes I\otimes I\otimes
X\otimes X)\Lambda_6(v_{15})(I\otimes I\otimes I\otimes I\otimes X\otimes X)\Lambda_5(X)\Lambda_2(X)(I\otimes
X\otimes I\otimes X\otimes I\otimes I)\Lambda_4(X)(I\otimes I\otimes X\otimes I\otimes I\otimes
I)\Lambda_3(X)(I\otimes I\otimes I\otimes I\otimes I\otimes X)\Lambda_3(X)(I\otimes I\otimes X\otimes I\otimes
I\otimes I)\Lambda_4(X)(I\otimes I\otimes I\otimes X\otimes I\otimes I)\Lambda_5(X)(I\otimes X\otimes I\otimes
I\otimes X\otimes I)\Lambda_2(X)\Lambda_6(v_{16})\Lambda_2(X)\Lambda_2(v_{17})(I\otimes X\otimes I\otimes
I\otimes X\otimes I)\Lambda_5(X)(I\otimes I\otimes I\otimes X\otimes I\otimes I)\Lambda_4(X)(I\otimes I\otimes
X\otimes I\otimes I\otimes I)\Lambda_3(X)(I\otimes X\otimes X\otimes X\otimes X\otimes
I)\Lambda_6(v_{18})(I\otimes X\otimes X\otimes I\otimes X\otimes I)\Lambda_6(X)(I\otimes I\otimes X\otimes
I\otimes X\otimes I)\Lambda_6(X)(I\otimes I\otimes I\otimes X\otimes I\otimes I)\Lambda_4(X)(I\otimes I\otimes
X\otimes I\otimes I\otimes I)\Lambda_3(X)(I\otimes I\otimes X\otimes I\otimes I\otimes I)\Lambda_4(X)(I\otimes
I\otimes I\otimes X\otimes I\otimes I)\Lambda_6(X)(I\otimes I\otimes X\otimes X\otimes X\otimes
X)\Lambda_3(X)\Lambda_5(X)(I\otimes I\otimes X\otimes X\otimes I\otimes X)$

        $  \Lambda_6(X)(I\otimes I\otimes I\otimes X\otimes I\otimes
X)\Lambda_4(X)\Lambda_5(X)(I\otimes X\otimes I\otimes I\otimes X\otimes I)\Lambda_2(X)(I\otimes X\otimes
I\otimes I\otimes X\otimes I)\Lambda_5(X)\Lambda_4(X)(I\otimes I\otimes I\otimes X\otimes I\otimes
X)\Lambda_6(X)(I\otimes X\otimes I\otimes X\otimes X\otimes X)\Lambda_2(X)(I\otimes X\otimes X\otimes I\otimes
I\otimes X)\Lambda_3(X)\Lambda_6(X)(I\otimes I\otimes I\otimes I\otimes I\otimes X)\Lambda_4(X)(I\otimes
X\otimes I\otimes X\otimes I\otimes I)\Lambda_2(X)(I\otimes X\otimes I\otimes X\otimes I\otimes
I)\Lambda_4(X)(I\otimes I\otimes I\otimes I\otimes I\otimes X)\Lambda_6(X)\Lambda_3(X)(I\otimes I\otimes
I\otimes I\otimes X\otimes X)\Lambda_3(X)\Lambda_5(X)(I\otimes X\otimes I\otimes I\otimes X\otimes
I)\Lambda_2(X) (I\otimes X\otimes I\otimes I\otimes X\otimes I)\Lambda_5(X)\Lambda_3(X)(I\otimes I\otimes
X\otimes I\otimes I\otimes X) \Lambda_6(X)(I\otimes I\otimes I\otimes I\otimes I\otimes X)\Lambda_5(X)(X\otimes
I\otimes I\otimes I\otimes X\otimes I) \Lambda_1(X)\Lambda_4(X)\Lambda_1(X)(X\otimes I\otimes I\otimes I\otimes
X\otimes I)\Lambda_5(X) (I\otimes I\otimes I\otimes I\otimes I\otimes X)\Lambda_6(X)(I\otimes X\otimes X\otimes
X\otimes X\otimes X)\Lambda_2(X) (I\otimes I\otimes I\otimes X\otimes I\otimes
I)\Lambda_4(X)\Lambda_3(X)\Lambda_4(X) (I\otimes I\otimes I\otimes X\otimes I\otimes I)\Lambda_2(X)(I\otimes
X\otimes I\otimes I\otimes I\otimes X)\Lambda_6(X) (I\otimes I\otimes I\otimes X\otimes I\otimes
X)\Lambda_4(X)(X\otimes I\otimes I\otimes I\otimes I\otimes I)\Lambda_1(X) \Lambda_3(X)\Lambda_1(X)(X\otimes
I\otimes I\otimes I\otimes I\otimes I)\Lambda_4(X) (I\otimes I\otimes I\otimes X\otimes I\otimes
X)\Lambda_6(X)(X\otimes I\otimes I\otimes I\otimes X\otimes X)\Lambda_1(X) \Lambda_5(X)\Lambda_1(X)(X\otimes
I\otimes I\otimes I\otimes I\otimes X)\Lambda_6(X) (I\otimes I\otimes I\otimes I\otimes I\otimes
X)\Lambda_5(X)(I\otimes I\otimes I\otimes X\otimes X\otimes I)\Lambda_4(X) (X\otimes I\otimes I\otimes I\otimes
I\otimes I)\Lambda_1(X)(X\otimes I\otimes I\otimes I\otimes I\otimes I)\Lambda_4(X) (I\otimes I\otimes I\otimes
X\otimes X\otimes I)\Lambda_5(X)(I\otimes I\otimes I\otimes I\otimes I\otimes X)\Lambda_6(X) (I\otimes X\otimes
I\otimes X\otimes X\otimes X)\Lambda_2(X)\Lambda_4(X)\Lambda_2(X) (I\otimes X\otimes I\otimes I\otimes I\otimes
X)\Lambda_6(X)(I\otimes I\otimes I\otimes I\otimes I\otimes X)\Lambda_4(X)(I\otimes X\otimes I\otimes X\otimes
I\otimes I)\Lambda_2(X)\Lambda_3(X)(X\otimes I\otimes X\otimes I\otimes I\otimes I) \Lambda_1(X)(X\otimes
I\otimes X\otimes I\otimes I\otimes I)\Lambda_3(X)\Lambda_2(X)(I\otimes X\otimes I\otimes X\otimes I\otimes
I)\Lambda_4(X)(I\otimes I\otimes I\otimes I\otimes I\otimes X)\Lambda_6(X)(I\otimes I\otimes I\otimes I\otimes
X\otimes X) \Lambda_3(X)(I\otimes I\otimes X\otimes I\otimes I\otimes I)\Lambda_5(X)(I\otimes X\otimes I\otimes
I\otimes X\otimes I)\Lambda_2(X)(X\otimes I\otimes I\otimes I\otimes I\otimes I)\Lambda_1(X)(X\otimes I\otimes
I\otimes I\otimes I\otimes I)\Lambda_2(X)(I\otimes X\otimes I\otimes I\otimes X\otimes I)\Lambda_5(X)(I\otimes
I\otimes X\otimes I\otimes I\otimes I)\Lambda_3(X)(I\otimes I\otimes I\otimes I\otimes I\otimes
X)\Lambda_6(X)(I\otimes I\otimes I\otimes I\otimes I\otimes X)\Lambda_5(X)(I\otimes I\otimes I\otimes I\otimes
X\otimes I)\Lambda_4(X)(I\otimes X\otimes I\otimes X\otimes I\otimes I)\Lambda_2(X)(X\otimes I\otimes I\otimes
I\otimes I\otimes I)\Lambda_1(X)(X\otimes I\otimes I\otimes I\otimes I\otimes I)\Lambda_2(X)(I\otimes X\otimes
I\otimes X\otimes I\otimes I)\Lambda_4(X)(I\otimes I\otimes I\otimes I\otimes X\otimes I)\Lambda_5(X)(I\otimes
I\otimes I\otimes I\otimes I\otimes X)\Lambda_6(X)(I\otimes X\otimes I\otimes X\otimes X\otimes
X)\Lambda_3(X)(I\otimes I\otimes I\otimes X\otimes I\otimes I)\Lambda_4(X)(I\otimes I\otimes I\otimes X\otimes
I\otimes I)\Lambda_3(X)(I\otimes I\otimes X\otimes I\otimes I\otimes X)\Lambda_6(X)(X\otimes I\otimes I\otimes
I\otimes I\otimes X)\Lambda_1(X)(I\otimes I\otimes X\otimes I\otimes I\otimes I)\Lambda_3(X)(I\otimes I\otimes
I\otimes X\otimes I\otimes I)\Lambda_4(X)(I\otimes I\otimes I\otimes X\otimes I\otimes I)\Lambda_3(X)(I\otimes
I\otimes X\otimes I\otimes I\otimes I)\Lambda_1(X)(X\otimes I\otimes I\otimes I\otimes I\otimes
X)\Lambda_6(X)(I\otimes I\otimes I\otimes X\otimes I\otimes X)\Lambda_4(X)(X\otimes I\otimes I\otimes X\otimes
I\otimes I)\Lambda_1(X)\Lambda_2(X)\Lambda_1(X)(X\otimes I\otimes I\otimes X\otimes I\otimes
I)\Lambda_4(X)(I\otimes I\otimes X\otimes X\otimes I\otimes I)\Lambda_3(X)(X\otimes I\otimes X\otimes I\otimes
I\otimes I)\Lambda_1(X)\Lambda_2(X)\Lambda_1(X)(X\otimes I\otimes X\otimes I\otimes I\otimes
I)\Lambda_3(X)(I\otimes I\otimes I\otimes X\otimes I\otimes I)\Lambda_3(X)(I\otimes I\otimes X\otimes I\otimes
I\otimes I)\Lambda_2(X)(X\otimes X\otimes I\otimes I\otimes I\otimes I)\Lambda_1(X)(X\otimes X\otimes I\otimes
I\otimes I\otimes I)\Lambda_2(X)(I\otimes I\otimes X\otimes I\otimes I\otimes I)\Lambda_3(X)(X\otimes X\otimes
I\otimes X\otimes I\otimes I)\Lambda_3(X)(I\otimes I\otimes I\otimes X\otimes I\otimes
I)\Lambda_3(X)\Lambda_4(X)(I\otimes I\otimes I\otimes X\otimes I\otimes I)\Lambda_3(X)(I\otimes X\otimes
X\otimes I\otimes I\otimes I)\Lambda_2(X)(I\otimes X\otimes X\otimes I\otimes I\otimes I)\Lambda_3(X)(I\otimes
I\otimes I\otimes X\otimes I\otimes I)\Lambda_4(X)(I\otimes X\otimes I\otimes X\otimes I\otimes
I)\Lambda_3(X)(I\otimes I\otimes X\otimes X\otimes I\otimes I)\Lambda_4(X)(I\otimes I\otimes X\otimes I\otimes
X\otimes X),$

 where
\begin{eqnarray*}
&&v_1=\left(%
\begin{array}{cc}
  \frac{1}{\sqrt{2}} & -\frac{1}{\sqrt{2}} \\
  -\frac{1}{\sqrt{2}} & -\frac{1}{\sqrt{2}} \\
\end{array}%
\right),~~
v_2=\left(%
\begin{array}{cc}
  \frac{1}{\sqrt{2}} & \frac{1}{\sqrt{2}} \\
  \frac{1}{\sqrt{2}} & -\frac{1}{\sqrt{2}} \\
\end{array}%
\right),~~v_3=\left(%
\begin{array}{cc}
  \frac{1}{\sqrt{3}} & \sqrt{\frac{2}{3}}\\
  \sqrt{\frac{2}{3}} & - \frac{1}{\sqrt{3}}\\
\end{array}%
\right),
~~v_4=\left(%
\begin{array}{cc}
  -\sqrt{\frac{2}{5}} & -\sqrt{\frac{3}{5}} \\
  -\sqrt{\frac{3}{5}} & \sqrt{\frac{2}{5}} \\
\end{array}%
\right),\\
&&v_5=\left(%
\begin{array}{cc}
  \frac{\sqrt{\frac{5}{2}}}{2} & -\frac{\sqrt{\frac{3}{2}}}{2} \\
  -\frac{\sqrt{\frac{3}{2}}}{2}& -\frac{\sqrt{\frac{5}{2}}}{2} \\
\end{array}%
\right),
~~v_6=\left(%
\begin{array}{cc}
  \frac{2}{\sqrt{11}} & -\sqrt{\frac{7}{11}} \\
  -\sqrt{\frac{7}{11}} & -\frac{2}{\sqrt{11}}\\
\end{array}%
\right),~~v_7=\left(%
\begin{array}{cc}
  \sqrt{\frac{11}{14}} & -\sqrt{\frac{3}{14}} \\
  -\sqrt{\frac{3}{14}} & -\sqrt{\frac{11}{14}} \\
\end{array}%
\right),
~~v_8=\left(%
\begin{array}{cc}
  2\sqrt{\frac{2}{11}} & -\sqrt{\frac{3}{11}} \\
  -\sqrt{\frac{3}{11}} & -2\sqrt{\frac{2}{11}} \\
\end{array}%
\right),\\
&&v_9=\left(%
\begin{array}{cc}
  \frac{3}{\sqrt{205}} & \frac{14}{\sqrt{205}} \\
  \frac{14}{\sqrt{205}} & - \frac{3}{\sqrt{205}} \\
\end{array}%
\right),~~v_{10}=\left(%
\begin{array}{cc}
  \sqrt{\frac{123}{131}} & -2\sqrt{\frac{2}{131}} \\
  -2\sqrt{\frac{2}{131}} & -\sqrt{\frac{123}{131}}\\
\end{array}%
\right),
~~v_{11}=\left(%
\begin{array}{cc}
  -\sqrt{\frac{131}{203}} & 6\sqrt{\frac{2}{203}} \\
  6\sqrt{\frac{2}{203}} & \sqrt{\frac{131}{203}}\\
\end{array}%
\right),~~v_{12}=\left(%
\begin{array}{cc}
  \frac{\sqrt{\frac{29}{2}}}{4} & -\frac{\sqrt{\frac{3}{2}}}{4} \\
  -\frac{\sqrt{\frac{3}{2}}}{4} & -\frac{\sqrt{\frac{29}{2}}}{4} \\
\end{array}%
\right),\\
&&v_{13}=\left(%
\begin{array}{cc}
  -5\sqrt{\frac{131}{5453}} & -33\sqrt{\frac{2}{5453}} \\
  -33\sqrt{\frac{2}{5453}} & 5\sqrt{\frac{131}{5453}} \\
\end{array}%
\right),~~v_{14}=\left(%
\begin{array}{cc}
  -\sqrt{\frac{1653}{1703}} & 5\sqrt{\frac{2}{1703}} \\
  5\sqrt{\frac{2}{1703}} & \sqrt{\frac{1653}{1703}}\\
\end{array}%
\right),~~v_{15}=\left(%
\begin{array}{cc}
  \sqrt{\frac{26}{29}} & \sqrt{\frac{3}{29}} \\
  \sqrt{\frac{3}{29}} & -\sqrt{\frac{26}{29}} \\
\end{array}%
\right),\\
&&v_{16}=\left(%
\begin{array}{cc}
  -\sqrt{\frac{13}{38}} & -\frac{5}{\sqrt{38}} \\
  -\frac{5}{\sqrt{38}} & \sqrt{\frac{13}{38}} \\
\end{array}%
\right),
v_{17}=\left(%
\begin{array}{cc}
  \frac{3}{\sqrt{13}} & \frac{2}{\sqrt{13}} \\
  \frac{2}{\sqrt{13}} & -\frac{3}{\sqrt{13}} \\
\end{array}%
\right),~~v_{18}=\left(%
\begin{array}{cc}
  -\frac{1}{\sqrt{5}} & \frac{2}{\sqrt{5}} \\
  \frac{2}{\sqrt{5}} & \frac{1}{\sqrt{5}} \\
\end{array}%
\right).
\end{eqnarray*}

 Barenco et al $^{16}$ showed that a set of gates consisting of all one-bit quantum gates ( of the form $C^0(u)$ )
 and the two-bit
controlled-not (CNOT) gates $C^1(X)=C^1\left (
\begin{array} {cc}
0&1\\
1&0\\
\end{array}\right )$ is universal in the sense that all unitary operations on
arbitrarily many bits can be expressed as compositions of these gates. They exhibited  a general simulation of
$C^n(u)$  for an arbitrary  one-bit unitary operation $u$ using only these basic gates.  Combining the results
obtained by Barenco et al and  the above decomposition of $U$ we can give the explicit construction of the
unitary operation $U$ using one-qubit gates and two-qubit CNOT gates. For saving space, we do not give an
implementation of $U$ in terms of one and two qubit operations and
 also do not depict out the quantum circuit illustrating the procedure of probabilistic
cloning quantum states.

In  summary, by means of the primitive operations consisting of single-qubit gates, two-qubit controlled-not
gates,  we construct an efficient quantum logic network for  probabilistic cloning the quantum states used in
implemented tasks for which cloning provides some enhancement in performance. We hope that this quantum logic
network will be realized by experiment.\\[0.1cm]

 {\noindent \bf\large ACKNOWLEDGEMENTS}\\[0.1cm]

 This work was supported  by National Natural Science Foundation of
China under Grant No. 10271081 and Hebei Natural Science Foundation under Grant No. 101094.\\[0.3cm]

{\noindent \small $^1$ W. K. Wootters and W. H. Zurek, "A single quantum cannot be cloned," Nature  {\bf
299}, 802-803 (1982).\\
$^2$ D. Dieks, "Communication by electron-paramagnetic-res devices," Phys. Lett. A {\bf 92},
271-272 (1982).\\
$^3$ V. Bu\v{z}ek and M. Hillery, "Quantum copying: beyond the no-cloning theorem," Phys. Rev. A {\bf
54}, 1844-1852 (1996).\\
$^4$ N. Gisin and S. Massar, "Optimal quantum cloning machines,"  Phys. Rev. Lett. {\bf 79}, 2153-2156 (1997).\\
$^5$ N. Gisin, "Quantum cloning without signalling," Phys. Lett. A {\bf 242}, 1-3 (1998).\\
$^6$ R. F. Werner, "Optimal cloning of pure states," Phys. Rev. A {\bf 58}, 1827-1832 (1998).\\
$^7$ M. Keyl and R. F. Werner, "Optimal cloning of pure states, testing single clones," J. Math. Phys. {\bf 40}, 3283-3299
  (1999).\\
$^8$ D. Bru${\rm \ss}$ and C. Macchiavello, "Optimal state estimation for d-dimensional quantum systems," Phys.
Lett. A {\bf 253}, 249-251 \\
$~~~$ (1999).\\
$^9$ D. Bru$\ss$, A. Ekert and C. Macchiavello, "Optimal universal quantum cloning and state estimation," Phys. Rev.
Lett. {\bf 81}, \\ $~~~~$ 2598-2601 (1998).\\
$^{10}$ V. Bu\v{z}ek, M. Hillery and R. Bednik, "Controlling the flow of quantum information in quantum cloners:
asymmetric \\$~~~~$ cloning," Acta Phys. Slov. {\bf 48}, 177-184 (1998).\\
$^{11}$ Nicolas J. Cerf, "Asymmetric quantum cloning in any dimension," J. Mod. Opt. {\bf 47}, 187-209 (2000).\\
$^{12}$ V. Bu\v{z}ek, S. Braunstein, M. Hillery and D. Bru${\rm \ss}$, "Quantum copying: a network," Phys. Rev.
A {\bf 56},
3446-3452 (1997).\\
$^{13}$ L. M. Duan and G. C. Guo, "Probabilistic cloning and identification of linearly independent quantum
states," Phys. Rev. \\
$~~~~$ Lett. {\bf 80},
4999-5002 (1998).\\
$^{14}$ E. F. Galv$\tilde{\rm a}$o and L. Hardy, "Cloning and quantum computation," Phys. Rev. A {\bf 62}, 022301 (2000).\\
$^{15}$ T. Gao, F. L. Yan and Z. X. Wang,  "Exact achievable efficiencies for probabilistically  cloning the states,"
 LANL e-print\\ $~~~~$ quant-ph/0307186.\\
$^{16}$ A. Barenco, C. H. Bennett, R. Cleve, D. P. DiVincenzo, N. Margolus, P. Shor, T. Sleator, J. A. Smolin,
and H. Weinfurter,\\ $~~~~$ "Elementary gates for quantum computation,"
 Phys. Rev. A {\bf 52}, 3457-3467 (1995).\\

\end{document}